\documentclass[useAMS,usenatbib]{mnras}

\usepackage[english]{babel}
\usepackage{amsmath,amssymb}
\usepackage{graphicx, subfig}
\usepackage[normalem]{ulem}

\usepackage{verbatim}

\usepackage{color}
\usepackage{multirow}
\usepackage{mathtools}
\usepackage{epstopdf}

\usepackage{url}
\usepackage{color}

\def\nat{Nature}

\def\mnras{MNRAS}
\def\apj{ApJ}
\def\apjl{ApJL}
\def\apjs{ApJS}
\def\aap{A\&A}

\def\be{\begin{equation}} 
\def\ee{\end{equation}} 
\def\ba{\begin{eqnarray}} 
\def\ea{\end{eqnarray}}

\def\msun{{\Msun}}

\def\HH{${\rm {H_2}}\,\,$}

\def\HeII{\hbox{He~$\scriptstyle\rm II\ $}}

\def\gsim{\lower.5ex\hbox{\gtsima}} 
\def\lsim{\lower.5ex\hbox{\ltsima}} \def\gtsima{$\; \buildrel > \over 
\sim \;$} \def\ltsima{$\; \buildrel < \over \sim \;$} \def\prosima{$\; 
\buildrel \propto \over \sim \;$} \def\gsim{\lower.5ex\hbox{\gtsima}} 
\def\lsim{\lower.5ex\hbox{\ltsima}} 
\def\simgt{\lower.5ex\hbox{\gtsima}} 
\def\simlt{\lower.5ex\hbox{\ltsima}} 
\def\simpr{\lower.5ex\hbox{\prosima}} \def\la{\lsim} \def\ga{\gsim} 
  
 \def\gtsima{$\; \buildrel > \over \sim \;$} 
\def\ltsima{$\; \buildrel < \over \sim \;$} 
\def\gsim{\lower.5ex\hbox{\gtsima}} 
\def\lsim{\lower.5ex\hbox{\ltsima}} 
\def\simgt{\lower.5ex\hbox{\gtsima}} 
\def\simlt{\lower.5ex\hbox{\ltsima}} 
\def\simpr{\lower.5ex\hbox{\prosima}} 
\def\la{\lsim} 
\def\ga{\gsim}

\def\msun{\,{\rm \Msun}}

\def\E3{{\cal E}_{\rm g}^{III}}

\def\r12{r_{1/2}} 
\def\x12{x_{1/2}} 
\def\v12{v_{1/2}}

\def\msun{{\rm M}_{\odot}}
\def\zsun{{\rm Z}_{\odot}}

\newcommand\textlcsc[1]{\textsc{\MakeLowercase{#1}}}
\newcommand{\quotes}[1]{``#1''}

\usepackage{verbatim}

\def\lya{{\rm Ly}\alpha}
\def\HEII{\hbox{He~$\scriptstyle\rm II$~}} 

\def\comp{\rm comp}
\def\pure{\rm pure}
\def\lum{{\rm erg}\,{\rm s}^{-1}}

\def\zcrit{Z_{\rm crit}}

\defcitealias{pallottini:2014_sim}{P14}
\defcitealias{Sobral:2015arXiv15}{S15}

\begin{document}

\date{}
\pagerange{\pageref{firstpage}--\pageref{lastpage}} \pubyear{2015}
%\title[Population III stars in LAEs]{Do Population III stars hide in Lyman-$\alpha$ emitters?}
%\title[The Brightest LAE: PopIII stars or Massive BH?]{The Brightest LAE: Pop III stars or Massive Black Hole?}
\title[The Brightest LAE: Pop III stars or BH?]{The Brightest Ly$\alpha$ Emitter: Pop III or Black Hole?}
\author[Pallottini et al.]{A. Pallottini$^{1}$\thanks{email: andrea.pallottini@sns.it}, A. Ferrara$^{1,2}$, F. Pacucci$^{1}$, S. Gallerani$^{1}$, S. Salvadori$^{3}$, R. Schneider$^{4}$,\newauthor D. Schaerer$^{5,6}$, D. Sobral$^{7,8,9}$, J. Matthee$^{9}$\\
$^{1}$Scuola Normale Superiore, Piazza dei Cavalieri 7, I-56126 Pisa, Italy\\
$^{2}$Kavli IPMU, The University of Tokyo, 5-1-5 Kashiwanoha, Kashiwa 277-8583, Japan\\
$^{3}$Kapteyn Astronomical Institute, Landleven 12, 9747 AD Groningen, The Netherlands\\
$^{4}$INAF, Osservatorio Astronomico di Roma, Via Frascati 33, 00040 Monteporzio Catone, Italy\\
$^5$ Observatoire de Gen\`{e}ve, Departement d'Astronomie, Universit\'{e} de Gen\`{e}ve, 51 Ch. des Maillettes, 1290 Versoix, Switzerland\\
$^6$ CNRS, IRAP, 14 Avenue E. Belin, 31400 Toulouse, France\\
$^7$ Instituto de Astrof\'{i}sica e Ci\^{e}ncias do Espa\c{c}o, Universidade de Lisboa, OAL, Tapada da Ajuda, PT1349-018 Lisboa, Portugal\\
$^8$ Departamento de F\'{i}sica, Faculdade de Ci\^{e}ncias, Universidade de Lisboa, Edif\'{i}cio C8, Campo Grande, PT1749-016 Lisbon, Portugal\\
$^9$ Leiden Observatory, Leiden University, P.O. Box 9513, NL-2300 RA Leiden, The Netherlands
}
\maketitle

\label{firstpage}

\begin{abstract}
CR7 is the brightest $z=6.6 \, \lya$ emitter (LAE) known to date, and spectroscopic follow-up by \citet[][]{Sobral:2015arXiv15} suggests that CR7 might host Population (Pop) III stars.
We examine this interpretation using cosmological hydrodynamical simulations. Several simulated galaxies show the same ``Pop III wave'' pattern observed in CR7. However, to reproduce the extreme CR7 $\lya$/HeII1640 line luminosities ($L_{\rm \alpha/He II}$) a top-heavy IMF and a massive ($\gsim 10^{7}\msun$) Pop III burst with age $\simlt 2$ Myr are required.
Assuming that the observed properties of $\lya$ and HeII emission are typical for Pop III, we predict that in the COSMOS/UDS/SA22 fields, 14 out of the 30 LAEs at $z=6.6$ with $L_{\alpha} >10^{43.3}\lum$ should also host Pop III stars producing an observable $L_{\rm He II}\gsim10^{42.7}\lum$.
As an alternate explanation, we explore the possibility that CR7 is instead powered by accretion onto a Direct Collapse Black Hole (DCBH). Our model predicts $L_{\alpha}$, $L_{\rm He II}$, and X-ray luminosities that are in agreement with the observations.
In any case, the observed properties of CR7 indicate that this galaxy is most likely powered by sources formed from pristine gas. We propose that further X-ray observations can distinguish between the two above scenarios.
\end{abstract}

\begin{keywords}
stars: Population III -- galaxies: high-redshift -- black hole physics
\end{keywords}

%%%%%%%%%%%%%%%%%%%%%%%
\section{Introduction}
%%%%%%%%%%%%%%%%%%%%%%%
The end of the Dark Ages is marked by the appearance of the first stars. Such -- Pop III -- stars had to form out of a pristine composition (H+He) gas with virtually no heavy elements. Lacking these cooling agents, the collapse had to rely on the inefficient radiative losses provided by \HH molecules. Mini-halos, i.e. nonlinear dark matter structures with mass $M_{h}\sim 10^{6-7}\msun$ collapsing at high redshift ($z\sim30$), are now thought to be the preferred sites of first star formation episodes \citep{Yoshida:2006ApJ,Turk:2009Sci,salvadori:2009mnras,greif:2012mnras,Visbal:2015arXiv}. Although the Initial Mass Function (IMF) of Pop III stars is largely uncertain, physical arguments suggest that they could have been more massive than present-day (Pop II) stars. Furthermore, the metals produced by Pop III stars polluted the surrounding gas \citep{Bromm:2002ApJ,wise:2012apj,xu:2013arxiv}, inducing a transition to the Pop II star formation mode (``chemical feedback'', \citealt[][]{schneider:2002apj,schneider:2006mnras}). Metal enrichment is far from being homogeneous, and pockets of pristine gas sustaining Pop III star formation can in principle persist down to $z\simeq 3-4$ \citep[][]{Tornatore:2007MNRAS,trenti:2009b,maio:2010mnras,salvadori:2014mnras,pallottini:2014_sim,Ma:2015MNRAS}, yielding Pop III star formation rate (SFR) densities of $\sim 10^{-4}\msun {\rm yr}^{-1} {\rm Mpc}^{-3}$, i.e. $\lsim 1$\% of the Pop II SFR density at those redshifts.

The search effort for Pop III stars at moderate and high redshifts has become increasingly intense in the last few years \citep[e.g.][]{kashikawa:2012ApJ,heap:2015arxiv}. Observationally, a galaxy hosting a recent ($t_{\star}\lsim 2\,{\rm Myr}$) Pop III star formation episode should show strong $\lya$ and \HEII lines and no metal lines \citep[e.g.][]{Schaerer:2002A&A,Raiter:2010A&A,kehrig:2015ApJ}. Until now, no indisputable evidence for Pop III stars in distant galaxies has been obtained, and observations have only yielded upper bounds on Pop III SFR \citep[e.g.][]{Cai:2011ApJ,Cassata:2013A&A,Zabl:2015arXiv15}. This situation might dramatically change following the recent observations of CR7 by \citet[][\citetalias{Sobral:2015arXiv15} hereafter]{Sobral:2015arXiv15}.

%\subsection{The case of CR7}\label{sec_cr7_case}
CR7 is the brightest $\lya$ emitter (LAE) at $z>6$, and it is found in the COSMOS field \citep[][]{Matthee:2015arXiv15}. Spectroscopic follow-up by \citetalias{Sobral:2015arXiv15} suggests that CR7 might host a PopIII-like stellar population. This is based on the astonishingly bright $\lya$ and \HeII lines ($L_\alpha\simeq10^{43.93}\lum$, $L_{\rm He II}\simeq10^{43.29}\lum$) and no detection of metal lines. \citetalias{Sobral:2015arXiv15} shows that CR7 can be described by a composite of a PopIII-like and a more normal stellar population, which would have to be physically separated, and that would be consistent with e.g. \citet{Tornatore:2007MNRAS}. HST imaging shows that CR7 is indeed composed of different components: 3 separate sub-systems (A, B, C) with projected separations of $\lsim 5 {\rm kpc}$. F110W(YJ) and F160W(H) band photometry indicates that clump A might be composed of young (blue) stars, while the stellar populations of B+C are old and relatively red.
The observed $\lya$ and \HEII lines are narrow (${\rm FWHM}\lsim 200\, {\rm km}\,{\rm s}^{-1}$ and ${\rm FWHM}\lsim 130\, {\rm km}\,{\rm s}^{-1}$, respectively), disfavoring the presence of an AGN or Wolf-Rayet (WR) stars, which are expected to produce much broader (${\rm FWHM}\gsim10^3 {\rm km}\,{\rm s}^{-1}$) lines \citep[e.g.][]{DeBreuck:2000AA,Brinchmann:2008MNRAS,Erb:2010ApJ}. \citetalias{Sobral:2015arXiv15} concluded that CR7 likely contains a \textit{composite} stellar population, with clump A being powered by a recent Pop III-like burst ($t_{\star}\lsim 2\,{\rm Myr}$), and clumps B+C containing an old ($t_{\star}\sim 350\,{\rm Myr}$) burst of Pop II stars with $M_{\star}\simeq10^{10}\msun$, largely dominating the stellar mass of the entire system.

Based on cosmological simulations that follows the simultaneous evolution of Pop II and Pop III stars \citep[][\citetalias{pallottini:2014_sim} hereafter]{pallottini:2014_sim}, we examine the interpretation of CR7 as a Pop III host system and explore its implications.
We also propose an alternate explanation, briefly discussed in \citetalias{Sobral:2015arXiv15}, where CR7 is powered by accretion onto a Direct Collapse Black Hole and suggest further tests.

%%%%%%%%%%%%%%%%
\section{Simulation overview}
%%%%%%%%%%%%%%%%
We use the $\Lambda$CDM cosmological\footnote{We assume a $\Lambda$CDM cosmology with total matter, vacuum and baryonic densities in units of the critical density $\Omega_{\Lambda}= 0.727$, $\Omega_{dm}= 0.228$, $\Omega_{b}= 0.045$, Hubble constant $\rm H_0=100~h~km~s^{-1}~Mpc^{-1}$ with $\rm h=0.704$, spectral index $n=0.967$, $\sigma_{8}=0.811$ \citep[][]{Larson:2011}.} hydrodynamical simulations presented in \citetalias{pallottini:2014_sim} (see that paper for a comprehensive description), obtained with a customized version of the Adaptive Mesh Refinement code \textlcsc{ramses} \citep{teyssier:2002} to evolve a ($10h^{-1}$ Mpc)$^{3}$ volume from $z=99$ to $z=4$, with a dark matter mass resolution of $\simeq 5\times 10^{5}\,h^{-1}\msun$, and an adaptive baryon spatial resolution ranging from $\simeq 20\,h^{-1}$~kpc to $\simeq 1\,h^{-1}$~kpc.
Star formation is included via sub-grid prescriptions based on a local density threshold. If the star forming cell gas has metallicity below (above) the critical metallicity, $\zcrit\equiv10^{-4}\zsun$, we label the newly formed stars as Pop III (Pop II). Supernova feedback accounts for metal-dependent stellar yields and return fractions appropriate for the relevant stellar population\footnote{While in \citetalias{pallottini:2014_sim} we explore different types of IMF for Pop III, here we show results assuming a Pop II-like Salpeter IMF. It is to note that our simulations suggests that the Pop III SFR seems almost independent from the IMF (see in particular Fig. 14 in \citetalias{pallottini:2014_sim}).}.
The simulated galaxy sample reproduces the observed cosmic SFR \citep[][]{Bouwens:2012ApJ,Zheng:2012Natur} and stellar mass density \citep[][]{Gonzalez:2011} evolution in the redshift range $4\leq z \lsim 10$, and -- as shown in \citet{pallottinicmb} -- P14 reproduce the observed the luminosity function at $z=6$. Additionally, the derived Pop III cosmic SFR density is consistent with current observational upper limits \citep[e.g.][]{Nagao2008A-Photometric-S,Cai:2011ApJ,Cassata:2013A&A}. To allow a direct comparison with CR7 we will concentrate on the analysis of the $z \simeq 6$ simulation output.

%-----------------------------------------------------------------
\subsection{Pop III-hosting galaxies}\label{sec_pop3wave}
%-----------------------------------------------------------------

\begin{figure}
\centering
\includegraphics[width=0.40\textwidth]{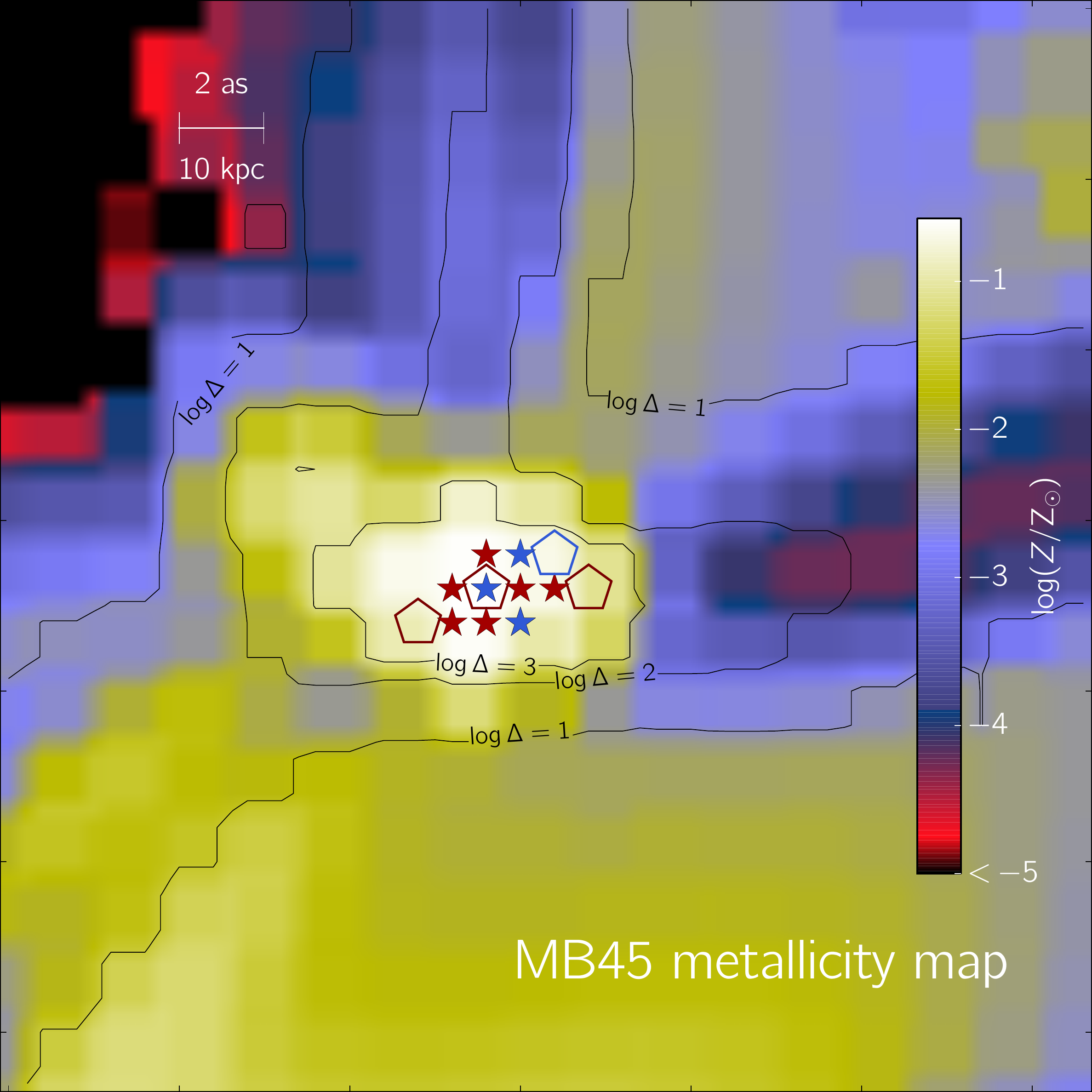}
\caption{Metallicity ($Z$) map centered on the simulated galaxy MB45. Pop II locations are marked with filled stars, for recent (blue, $t_{\star}\lsim 2~{\rm Myr}$) and old (red) formation events, respectively. Locations of the recent (old) Pop III burst are shown by the open blue (red) pentagons. Overdensity ($\log\Delta=1,\, 2, {\rm and} \, 3$) contours are plotted with black lines. The scale is indicated in arcsecond and kpc.
\label{fig_mappa_mb45}}
\end{figure}

As noted in \citetalias{Sobral:2015arXiv15}, the interpretation of CR7 fits in the \quotes{Pop III wave} scenario suggested by \citet[][]{Tornatore:2007MNRAS}. As an example of Pop III wave in action, we show the case of \quotes{MB45}, a simulated \citetalias{pallottini:2014_sim} galaxy with total stellar mass $M_{\star} = 10^{7.9}\msun$. In Fig. \ref{fig_mappa_mb45}, we plot the metallicity ($Z$) and overdensity ($\Delta$) map around MB45. The star formation history in MB45 starts with a Pop III event. These stars explode as supernovae enriching with metals the central regions of MB45. As a result, star formation there continues in the Pop II mode, while in the less dense external regions, not yet reached by the metal-bearing shocks, Pop III stars can still form. The process repeats until the unpolluted regions have densities exceedingly low to sustain star formation.

The total (i.e. old+young stars) Pop III mass in MB45 is $M_{3}\simeq 10^{6.8}\msun$; about 20\% of this stellar mass formed in a recent burst (age $t_{\star}\lsim 2~{\rm Myr}$). The total stellar mass ($M_\star\simeq 10^8 M_{\odot}$) of MB45 is dominated by Pop II stars produced at a rate ${\rm SFR}_{2}\simeq 0.5 \, \msun\,{\rm yr}^{-1}$. Thus, while MB45 formation activity proceeds in the Pop III wave mode and resembles that of CR7, the physical properties of MB45 and CR7 are different, because of the 2 orders of magnitude difference in total stellar mass. A direct comparison between the PopIII-PopII separation in CR7 (projected $\simeq5\,{\rm kpc}$) and MB45 (10 kpc, projected $\lsim5\,{\rm kpc}$), although fairly consistent, might not be very meaningful due to the  different mass of the two systems. This is because the separation depends on the mass-dependent metallicity profile in galaxy groups (see Fig.1 in \citealt{pallottini:2014_cgmh}). However, we note that our current simulated volume is simply to small to be able to directly recover sources such as CR7, with volume densities of $\sim10^{-6}$\,Mpc$^{-3}$.

\begin{figure}
\centering
\includegraphics[width=0.47\textwidth]{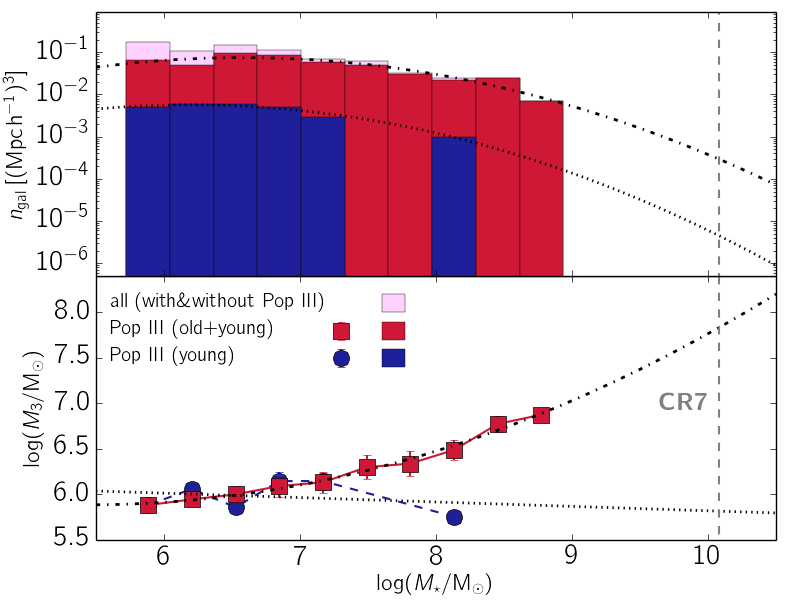}
\caption{{\bf Upper panel}: Number density ($n_{gal}$) of simulated galaxies as a function of their total stellar mass ($M_{\star} = M_{2} + M_{3}$) for: (i) all galaxies (pink bar), (ii) galaxies with old+young Pop III stars (red bars), (iii) galaxies with a young Pop III component (blue bars). All quantities are averaged on $\log(\Delta{M} /\msun)\simeq 0.3$ bins. Black dot-dashed and dotted lines correspond to the best-fit analytical extrapolation of $n_{gal}$.
{\bf Lower panel}: Mass of Pop III ($M_{3}$) in composite galaxies considering young (blue circle) and old+young (red squares) stars. In both panels the vertical dashed line indicates the value of $M_{\star}$ for CR7 as inferred by SED fitting (see Sec. \ref{sez_emission_helioealpha}).
\label{fig_fractionandpdf}}
\end{figure}

The simulated volume contains many galaxies hosting Pop III star formation, whose number density ($n_{gal}$) as a function of their total stellar mass ($M_{\star}$) is shown in Fig. \ref{fig_fractionandpdf}, along with their Pop III mass ($M_{3}$). Pop III stars are found preferentially in low-mass systems $M_{\star}\lsim 10^{6.5}\msun$, that typically form these stars in a series of a few $M_{3}\simeq 10^6\msun$ bursts\footnote{We refer to App. A of \citetalias{pallottini:2014_sim} for possible resolution effects.}, before Pop III formation is quenched by chemical feedback. 
 
Note that all larger galaxies, $M_{\star}\gsim 10^7\msun$, contain some Pop III component inherited from progenitor halos. We can then regard $M_{3}$ calculated by summing old and young Pop III stars as a solid upper bound to the total Pop III mass produced during the galaxy lifetime. The size of the simulation volume, dictated by the need of resolving the very first star-forming halos, is too small to fairly sample the mass function of galaxies with $M_{\star}\gsim 10^{9}\msun$ and more massive. To make predictions in this high-mass range, we slightly extrapolate the simulated trends of $n_{gal}$ and $M_{3}$ (lines in Fig. \ref{fig_fractionandpdf}). We caution that such extrapolation might imply uncertainties.

%

%-----------------------------------------------------------------------
\subsection{Ly$\mathbf{\alpha}$ and \HEII emission}\label{sez_emission_helioealpha}
%-----------------------------------------------------------------------

We label as {\it pure} ({\it composite}) galaxies whose emission is produced by Pop II only (Pop II+Pop III) stars. For a pure galaxy, the $\lya$ line luminosity is \citep[e.g.][]{dayal:2008mnras}:
\begin{subequations}\label{eqs_luminosita}
\be\label{eq_lum_ha_pure}
L_{\alpha}^{\pure}=2.8\times 10^{42} A_{\alpha}\, {\rm SFR}_{2} \,\, \lum\, ,
\ee
where $A_{\alpha}$ is an attenuation factor that accounts for both internal (interstellar medium) absorption and intergalactic medium transmissivity; Pop II SFR is expressed in $\msun$yr$^{-1}$ and to relate this to the $M_{\star}$ we assume ${\rm SFR}_2/{M_{\star}}={\rm sSFR}\simeq2.5\, {\rm Gyr}^{-1}$, consistent with our simulations and observations \citep[][see Sec. \ref{sec_predictions}]{Daddi:2007ApJ,McLure:2011MNRAS,Gonzalez:2011}.
Pure galaxies have no \HEII emission. For a composite galaxy, the $\lya$ emission is the sum of the contributions from Pop II and Pop III components:
\be\label{eq_lum_ha}
L_{\alpha}^{\comp}= L_{\alpha}^{\pure} + A_{\alpha} l^{\alpha}_{3}M_{3} \,,
\ee
where $l^{\alpha}_{3}$ is the Pop III $\lya$ line luminosity per unit stellar mass\footnote{We are implicitly assuming that Pop III stars form in a burst, an assumption justified by the analysis presented in Sec. \ref{sec_pop3wave}.}. Analogously, the \HEII emission is given by
\be\label{eq_lum_he}
L_{\rm He II}= l^{\rm He II}_{3}M_{3}\,,
\ee
\end{subequations}
where $l^{\rm He II}_{3}$ is the Pop III \HEII luminosity per unit stellar mass. Both $l^{\alpha}_{3}$ and $l^{\rm He II}_{3}$ depend on the IMF and burst age, $t_{\star}$. We adopt the Pop III models by \citet[][]{Schaerer:2002A&A} and \citet[][]{Raiter:2010A&A}, and we use a Salpeter IMF (power-law slope $\alpha=-2.35$), with variable lower ($m_{low}$) and upper ($m_{up}$) limits. As long as $m_{up}\simgt 10^2 \msun$, the results are very weakly dependent on the upper limit, which we therefore fix to $m_{up}=10^3 \msun$, leaving $m_{low}$ as the only free-parameter.

As noted by \citetalias{Sobral:2015arXiv15}, to reproduce CR7 $L_{\alpha}$ and $L_{\rm He II}$ with Pop IIIs, a mass of $M_{3}\simeq 10^{7 - 9}\msun$ \textit{newly-born} ($t_{\star} \simlt 2-5$ Myr) stars is required, depending on the IMF. Such a large amount of \textit{young} Pop III stars is contained in none of \citetalias{pallottini:2014_sim} galaxies and it is not predicted by the adopted analytical extrapolation (see Fig. \ref{fig_fractionandpdf}). Thus none of the simulated composite galaxies would reproduce CR7 line emission. However, it is possible that CR7 might have experienced a more vigorous Pop III star formation burst as a result of a very rare event -- e.g. a recent major merger -- not frequent enough to be captured in our limited box volume. As an estimate we adopt the value of $M_{3}$ resulting from the sum of all (old+young) Pop III stars formed in our galaxies. 
%This is equivalent to say that each galaxy must have formed all its Pop III stars in a single burst within $\simlt 2$ Myr.

Under this hypothesis, we can fix $m_{low}$, by using the zero age main sequence (ZAMS) tracks ($t_{\star}=0$). By SED fitting, \citetalias{Sobral:2015arXiv15} shows that CR7 Pop II stellar mass (completely contained in clumps B+C) is likely $M_\star\simeq M_{2}\simeq10^{10}\msun$. From the lower panel of Fig. \ref{fig_fractionandpdf}, we find that this mass corresponds to a Pop III mass of $M_{3} \sim 10^{7.5}\msun$. 
From Pop III SED fits of region A \citetalias{Sobral:2015arXiv15} estimate $M_{3} \sim 10^{7}\msun$. As these stars must be located in CR7 clump A, whose \HeII luminosity is $L_{\rm He II}=10^{43.3}\lum$, eq. \ref{eq_lum_he} requires that $l^{\rm He II}_{3}=10^{35.5}\lum\,\msun^{-1}$. In turn this entails a top-heavy IMF with $m_{low} = 6.7\, \msun$.

Having fixed the IMF, we can readily derive the predicted Pop III contribution to the $\lya$ emission; this turns out to be $l^{\alpha}_{3}=10^{36.7}\lum\,\msun^{-1}$. CR7 has an observed $L_{\alpha}= 10^{43.9}\lum$, with no contribution from clumps B+C (${\rm SFR}_{2}=0$). This comparison allows us to determine, using eq. \ref{eq_lum_ha}, the Ly$\alpha$ line attenuation factor, $A_{\alpha}=10^{-0.57}$. 

Roughly 66\% of the line luminosity is therefore damped, a figure consistent with other derivations \citep[e.g.][]{dayal:2008mnras},
and with the analysis of \citetalias{Sobral:2015arXiv15}. The above procedure provides a basis to model $\lya$ and \HEII emission for both pure and composite galaxies, assuming that the properties of the Pop III component are similar to those derived from CR7.

%%%%%%%%%%%%%%%%%%%%%%%
\section{Predictions for bright LAEs}\label{sec_predictions}
%%%%%%%%%%%%%%%%%%%%%%%
\begin{figure}
\centering
\includegraphics[width=0.47\textwidth]{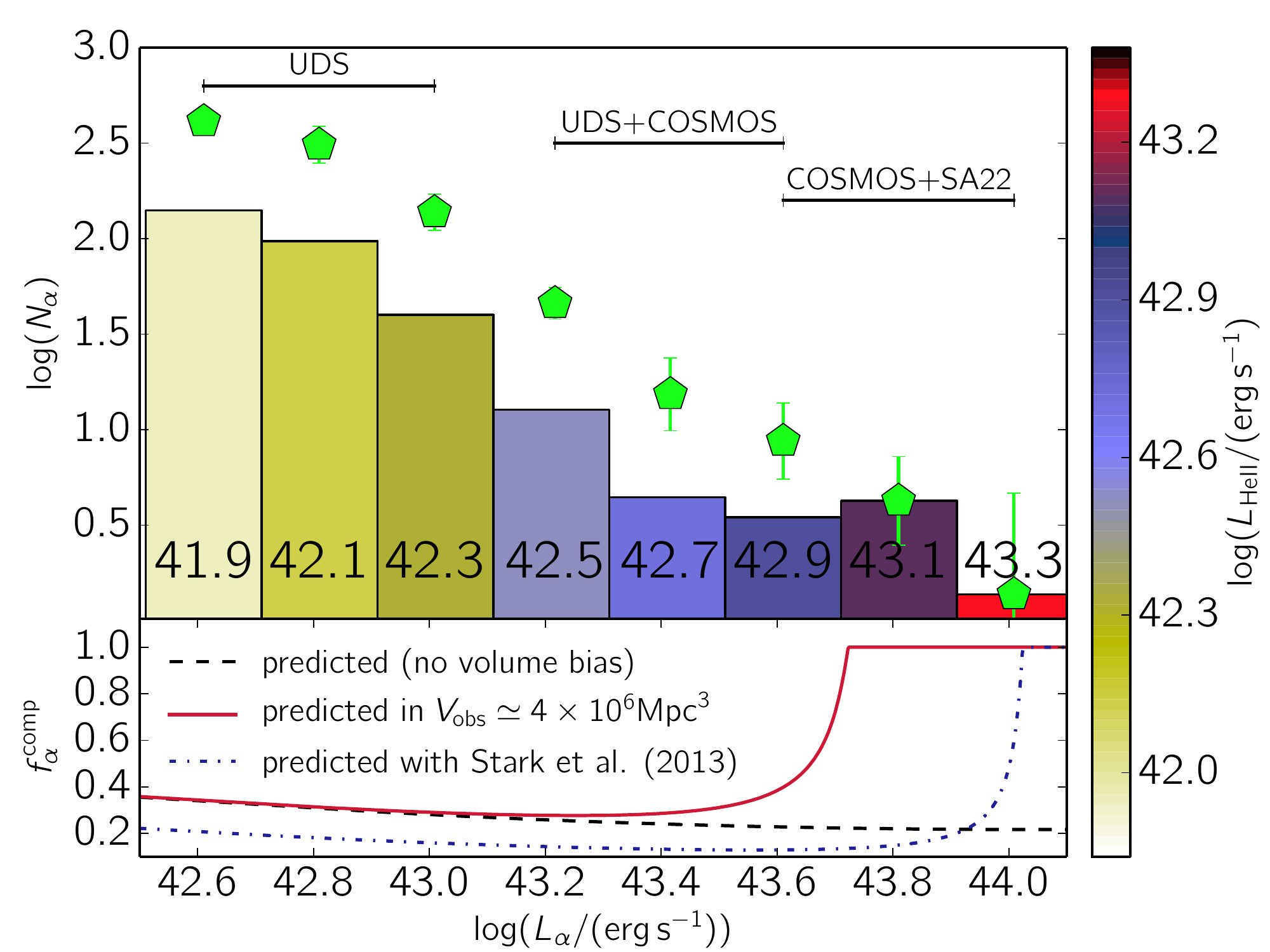}
\caption{
{\bf Upper panel}: Number of LAEs ($N_{\alpha}$) as a function of their $\lya$ luminosity ($L_{\alpha}$) in the COSMOS/UDS/SA22 fields \citep[][green pentagons]{Matthee:2015arXiv15}. The histogram shows the predicted number of composite (i.e. Pop III hosts) LAEs in the same survey; the bar colors indicate the expected $L_{\rm He II}$ emission, also shown by the numbers in the bars. 
{\bf Lower panel}:
Fraction of composite galaxies ($f_{\alpha}^{\rm comp}$, eq. \ref{cfrac}) for which the correction for the finite volume bias (eq. \ref{eq_bias_fcomp}) of \citet{Matthee:2015arXiv15} survey has been included (red solid) or neglected (black dashed).
\label{fig_alphaluminosity}}
\end{figure}

Starting from the assumption that CR7 is a \quotes{typical} composite galaxy, and using $m_{low} = 6.7\,\msun$ and $A_{\alpha}=10^{-0.57}$, we can now predict how many LAEs among those observed by \citet{Matthee:2015arXiv15} are composite galaxies, i.e. contain Pop III stars. The number of LAEs in the COSMOS/UDS/SA22 fields with luminosity $L_\alpha$ is $N_{\alpha}=\Phi_{\alpha}(L_\alpha)\,V_{obs}$, where $\Phi_{\alpha}$ is the observed $\lya$ luminosity function and $V_{obs}=4.26\times10^6 \, {\rm Mpc}^{3}$ is the observed volume. Among these, a fraction
\begin{subequations}
\be
f^{\comp}_{\alpha} = N_{\comp}/(N_{\comp}+N_{\pure})
\label{cfrac}
\ee
contain Pop III stars, where $N_{\comp}$ ($N_{\pure}$) is the number of composite (pure) galaxies in $V_{obs}$ at a given $L_\alpha$.

Eq.s \ref{eq_lum_ha_pure} and \ref{eq_lum_ha}, show that a given $L_{\alpha}$ can be produced by a composite galaxy with a lower $M_{\star}$ with respect to a pure (PopII) galaxy. For instance, $L_{\alpha}\simeq10^{43.5}\lum$ requires $M_{\star}\simeq10^{10.5}\msun$ for a pure galaxy, but only $M_{\star}\simlt 10^{9.5}\msun$ for a composite one. Such large objects are very rare at the redshift of CR7 ($z=6.6$) and in the observed volume\footnote{As a reference, a $M_{\star}\sim10^{10.5}\msun$ is hosted in a dark matter halo of mass $M_{h}\sim10^{12.5}\msun$, whose abundance is $n_h\sim10^{-6}{\rm Mpc}^{-3}$ at $z\simeq6$ \citep[e.g.][]{sheth:1999MNRAS}.}. Therefore, it is important to account for the statistical (Poisson) fluctuations of the galaxy number counts as follows:
\be\label{eq_bias_fcomp}
N_{\kappa}=n_{\kappa} V_{obs} (1 \pm (n_{\kappa}V_{obs})^{-1/2})\,,
\ee
\end{subequations}
where $\kappa =$~(composite, pure). The distribution of $n_{\comp}(M_\star)$ is shown in the upper panel of Fig. \ref{fig_fractionandpdf} as the \quotes{young} Pop III curve\footnote{As noted in Sec. \ref{sec_pop3wave}, all galaxies with $M_{\star}>10^7$ have old Pop III stars, thus considering the \quotes{old+young} track for $n_{\comp}$ would yield an unrealistically high composite number.}; while $n_{\pure}$ accounts for the remaining galaxies. The effect of the finite volume effects on $f^{\comp}_{\alpha}$ can be appreciated from the lower panel of Fig. \ref{fig_alphaluminosity}. Assuming a higher ${\rm sSFR}=5\,{\rm Gyr}^{-1}$ \citep[e.g.][]{Stark:2013ApJ} yields the modifications shown by the dash-dotted line.

In the upper panel of Fig. \ref{fig_alphaluminosity} we plot the LAE number ($N_{\alpha}$) as a function of $L_{\alpha}$. \citet{Matthee:2015arXiv15} observations (green pentagons) are shown along with our predictions for the composite LAE and expected $L_{\rm He II}$ emission. CR7 is the most luminous LAE observed, and it is in the brightest luminosity bin ($L_{\alpha}=10^{44\pm0.1}\lum$); by assumption CR7 is a composite galaxy. If so, we then predict that out of the $46$ ($30$) LAEs with $L_{\alpha}=10^{43.2\pm0.1}\lum$ ($>10^{43.3}\lum$, cumulative), $13$ ($14$) must also be composite galaxies\footnote{The predicted number would become $7$ ($7$), by assuming sSFR from \citet[][]{Stark:2013ApJ}.}, with observable $L_{\rm He II}\simeq10^{42.5}\lum$ ($\gsim10^{42.7}\lum$). Follow-up spectroscopy of those luminous Lyman-$\alpha$ emitters at $z=6.6$ will allow to test this prediction. We recall that this test assumes that all Pop III give raise to the same Ly$\alpha$ and \HEII emission as inferred from CR7, that requires that all the Pop III stellar mass was formed in a single burst with age $\lsim2$~Myr.

Particularly in the regime where $f_{\rm comp} < 1$ (see lower panel in Fig. \ref{fig_alphaluminosity}), a sample of LAEs is needed to test our model predictions. For example for \quotes{Himiko}, the second most luminous\footnote{As re-computed in \citetalias{Sobral:2015arXiv15} using Y band to estimate the continuum, in order to match the calculation for CR7.} confirmed LAE at $z = 6.6$ with $L_{\alpha}\simeq10^{43.4}\lum$ \citep[][]{Ouchi:2009ApJ}, for which recent VLT/X-Shooter observations have provided a $3 \sigma$ limit of $L_{\rm He II}\lsim10^{42.1}\lum$ \citep{Zabl:2015arXiv15}, our model predicts $L_{\rm He II}\simeq10^{42.7}\lum$ i.e. a four times higher \HEII~luminosity. However, this is predicted only for $f_{\rm comp} \simeq 20-30\%$ of galaxies at this $L_\alpha$.

%%%%%%%%%%%%%%%%%%%%%%%%%%%%%%%%%%%%%%%%%%%%%%
\section{Alternative interpretation}\label{sec:alternatives}
%%%%%%%%%%%%%%%%%%%%%%%%%%%%%%%%%%%%%%%%%%%%%%

Given the extreme conditions required to explain the observed properties of CR7 in terms of Pop III stars and a set of assumptions, it is worth exploring alternative interpretations. The most appealing one involves Direct Collapse Black Holes (DCBH), which is briefly discussed in \citetalias{Sobral:2015arXiv15}. High-$z$ pristine, atomic halos ($M_{h} \simgt 10^{8}\msun$) primarily cool via $\lya$ line emission. In the presence of an intense Lyman-Werner (LW, $E_{\gamma}=11.2-13.6 \, \mathrm{eV}$) irradiation, \HH molecule photo-dissociation enforces an isothermal collapse \citep{Shang_2010,Latif_2013c,Agarwal:2013MNRAS,yue:2014mnras}, finally leading to the formation of a DCBH of initial mass $M_\bullet \simeq 10^{4.5-5.5} \msun$ \citep{Begelman_2006, Volonteri_2008, Ferrara_2014}, eventually growing up to $10^{6-7} \msun$ by accretion of the halo leftover gas.

In CR7, clump A appears to be pristine, and it is irradiated by a LW flux from B+C\footnote{The LW is estimated by accounting for the stellar properties of clump B+C (in particular see Fig. 8 in \citetalias{Sobral:2015arXiv15}), and by assuming a $5$~kpc distance between B+C and A.} of $\sim 5\times 10^{-18} \lum\, {\rm cm}^{-2}{\rm Hz}^{-1}\, {\rm sr}^{-1}$, well in excess of the required threshold for DCBH formation \citep{Shang_2010,Latif_2013c,Sugimura_2014,Regan_2014}. Thus CR7 might be a perfect host for a DCBH.

We investigate the time-evolving spectrum of an accreting DCBH of initial mass $M_{\bullet} = 10^5 \msun$ by coupling a 1D radiation-hydrodynamic code \citep{Pacucci_2015} to the spectral synthesis code \textlcsc{cloudy} \citep{cloudy:2013}, as detailed in \citet{pacucci_2015_spettro}. The DCBH intrinsic spectrum is taken from \citet[][]{Yue_2013}. The DCBH is at the center of a halo of total gas mass $M_{g} \simeq 10^7 \msun$, distributed with a core plus a $r^{-2}$ density profile spanning up to 10 pc. The accretion is followed until complete depletion of the halo gas, i.e. for $\simeq 120 \, {\rm Myr}$. During this period the total absorbing column density of the gas varies from an initial value of $\simeq 3.5 \times 10^{24} {\rm cm}^{-2}$ to a final value $\ll 10^{22} {\rm cm}^{-2}$, i.e. from mildly Compton-thick to strongly Compton-thin. Note that while $\lya$ attenuation by the interstellar medium is included, we do not account for the likely sub-dominant IGM analogous effect.

\begin{figure}
\centering
\includegraphics[angle=0,width=0.49\textwidth]{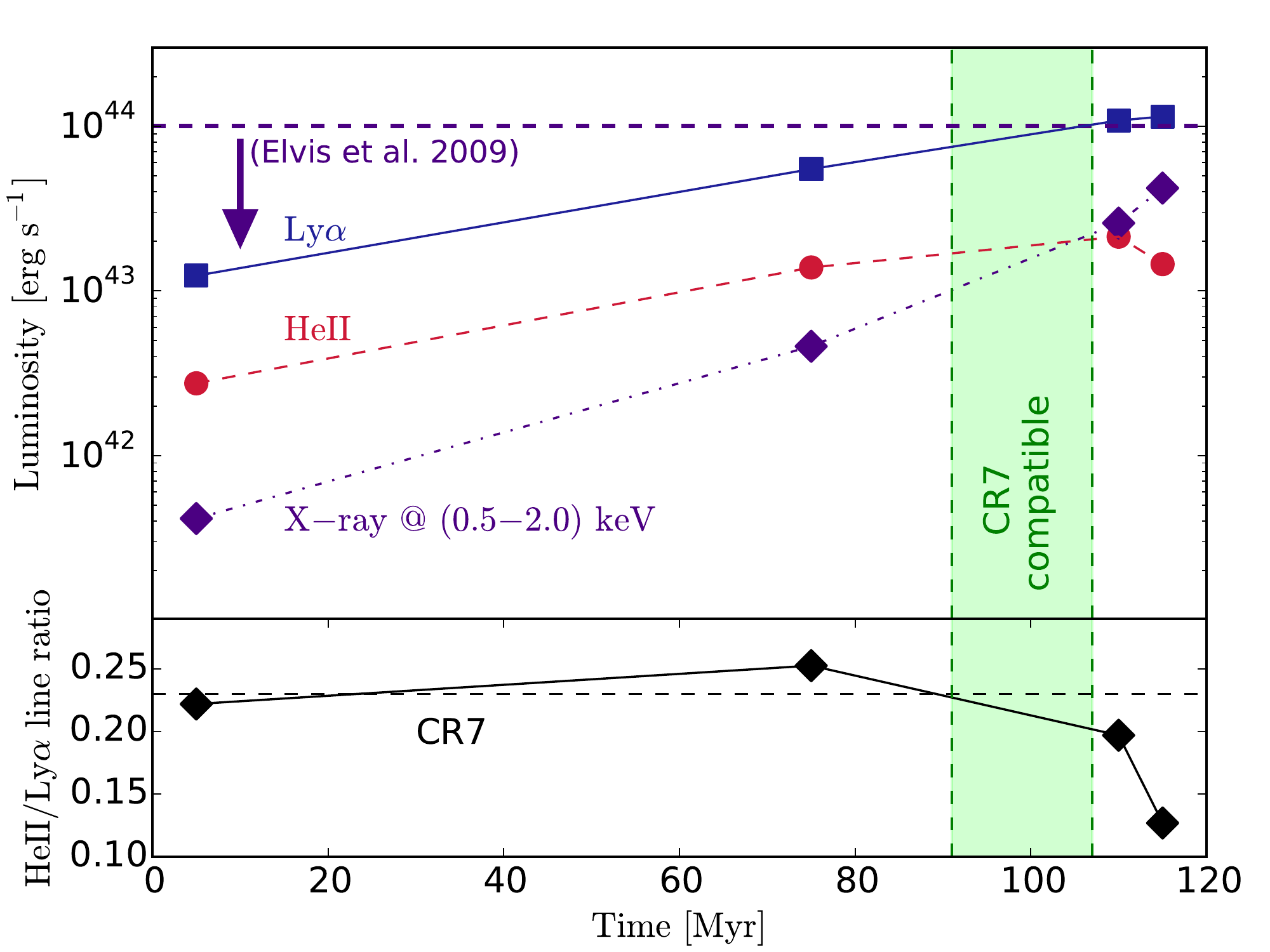}
\caption{
{\bf Upper panel}: Time evolution of the $\lya$ (blue solid line), \HEII (red dashed line) line and X-ray (violet dot dashed line) luminosities calculated for the accretion process onto a DCBH of initial mass $10^5 \msun$. The green shaded region indicates the period of time during which our simulations are compatible with CR7 observations. The current upper limit for X-ray is $\lsim 10^{44} \, \lum$, \citep[][horizontal violet line]{Elvis_2009}.
{\bf Lower panel}:
Time evolution of the ${\rm He II}/\lya$ lines ratio. The black horizontal dashed line indicates the observed values for CR7.
\label{fig_dcbh}}
\end{figure}

Fig. \ref{fig_dcbh} shows the time evolution of the $\lya$, \HEII and X-ray (0.5-2 keV) luminosities. Both $\lya$ and \HEII are consistent with the observed CR7 values during an evolutionary phase lasting $\simeq 17\, {\rm Myr}$ (14\% of the system lifetime), longer than the shorter period ($t_{\star}\simlt 2$~Myr) of our assumption for a massive Pop III burst.

The equivalent width of the \HeII line in the CR7 compatibility region ranges from 75 to 85$\,$ A. The column density during the CR7-compatible period is $\simeq 1.7 \times 10^{24}\mathrm{cm^{-2}}$, i.e. mildly Compton-thick. The associated X-ray luminosity is $\lsim 10^{43}\lum$, fully consistent with the current upper limit for CR7 ($\lsim 10^{44}\lum$, \citealt{Elvis_2009}). Deeper X-ray observations of CR7 might then confirm the presence of the DCBH. However, this limit is already obtained with 180\,ks of integration time on Chandra, meaning that a stringent test might only be possible with the next generation of X-ray telescopes.

%%%%%%%%%%%%%%%%%%%%%%
\section{Conclusions}
%%%%%%%%%%%%%%%%%%%%%%

CR7 is the brightest $z=6.6$ LAE in the COSMOS field \citep[][]{Matthee:2015arXiv15}. Spectroscopic follow-up \citep[][]{Sobral:2015arXiv15} suggests that CR7 might host Pop III stars, along with Pop II and thus be explained by a \quotes{Pop III wave} scenario. We have further investigated such interpretation using cosmological simulations following the formation of Pop II and Pop III stars in early galaxies.

We find simulated galaxies (like MB45 in Fig. \ref{fig_mappa_mb45}) hosting both Pop III and Pop II stars at $z=6.0$. Such \quotes{composite} galaxies have morphologies similar to that of CR7 and consistent with the \quotes{Pop III wave scenario}. However, to reproduce the extreme CR7 $\lya$/HeII1640 line luminosities, a top-heavy IMF combined with a massive ($M_{3} \gsim 10^{7}\msun$) Pop III burst of young stars ($t_{\star}\simlt 2-5$ Myr) is required. Our simulations do not predict such large burst, i.e. $M_{3}\simeq10^{6}\msun$, but our volume is also smaller than that used to discover CR7. Nonetheless, assuming that CR7 is typical of all metal-free components in our simulations, we predict that in the combined COSMOS, UDS and SA22 fields, out of the 30 LAEs with $L_{\alpha} >10^{43.3}\lum$, 14 should also host Pop III stars producing an observable $L_{\rm He II}\gsim10^{42.7}\lum$.

Given the extreme requirements set by the Pop III interpretation, we explored the possibility that CR7 is instead powered by accretion onto a Direct Collapse Black Hole (DCBH) of initial mass $10^5 \msun$. The predicted $L_{\alpha}$ and $L_{\rm He II}$ match CR7 observations during a time interval of $\sim 17\, {\rm Myr}$ ($\sim 14\%$ of the system lifetime). The predicted CR7 luminosity at 0.5-2 keV, $\lsim 10^{43} \, \lum$, is significantly below the current upper limit, i.e. $\lsim 10^{44} \, \lum$.

We conclude that the DCBH interpretation of CR7 is very appealing, and competitive with the explanation involving a massive Pop III burst.
For both explanations, the dominant ionizing source of this galaxy should have formed from pristine gas. Deep X-ray observations and other follow-up observations should allow to shed more light on this very peculiar source.

\section*{Acknowledgments}
SS acknowledges support from the Netherlands Organization for Scientific research (NWO), VENI grant 639.041.233. RS acknowledges support from the European Research Council under the European Union (FP/2007-2013)/ERC Grant Agreement n. 306476. D. Sobral acknowledges (i) financial support from the NWO through a Veni fellowship and (ii) funding from FCT through a FCT Investigator Starting Grant and Start-up Grant (IF/01154/2012/CP0189/CT0010) and from FCT grant PEst-OE/FIS/UI2751/2014.

\bibliographystyle{mnras}
\bibliography{master}

\begin{thebibliography}{}
\makeatletter
\relax
\def\mn@urlcharsother{\let\do\@makeother \do\$\do\&\do\#\do\^\do\_\do\%\do\~}
\def\mn@doi{\begingroup\mn@urlcharsother \@ifnextchar [ {\mn@doi@}
  {\mn@doi@[]}}
\def\mn@doi@[#1]#2{\def\@tempa{#1}\ifx\@tempa\@empty \href
  {http://dx.doi.org/#2} {doi:#2}\else \href {http://dx.doi.org/#2} {#1}\fi
  \endgroup}
\def\mn@eprint#1#2{\mn@eprint@#1:#2::\@nil}
\def\mn@eprint@arXiv#1{\href {http://arxiv.org/abs/#1} {{\tt arXiv:#1}}}
\def\mn@eprint@dblp#1{\href {http://dblp.uni-trier.de/rec/bibtex/#1.xml}
  {dblp:#1}}
\def\mn@eprint@#1:#2:#3:#4\@nil{\def\@tempa {#1}\def\@tempb {#2}\def\@tempc
  {#3}\ifx \@tempc \@empty \let \@tempc \@tempb \let \@tempb \@tempa \fi \ifx
  \@tempb \@empty \def\@tempb {arXiv}\fi \@ifundefined
  {mn@eprint@\@tempb}{\@tempb:\@tempc}{\expandafter \expandafter \csname
  mn@eprint@\@tempb\endcsname \expandafter{\@tempc}}}

\bibitem[\protect\citeauthoryear{{Agarwal}, {Davis}, {Khochfar}, {Natarajan}
  \& {Dunlop}}{{Agarwal} et~al.}{2013}]{Agarwal:2013MNRAS}
{Agarwal} B.,  {Davis} A.~J.,  {Khochfar} S.,  {Natarajan} P.,   {Dunlop}
  J.~S.,  2013, \mn@doi [\mnras] {10.1093/mnras/stt696}, \href
  {http://adsabs.harvard.edu/abs/2013MNRAS.432.3438A} {432, 3438}

\bibitem[\protect\citeauthoryear{{Begelman}, {Volonteri}  \& {Rees}}{{Begelman}
  et~al.}{2006}]{Begelman_2006}
{Begelman} M.~C.,  {Volonteri} M.,   {Rees} M.~J.,  2006, \mn@doi [\mnras]
  {10.1111/j.1365-2966.2006.10467.x}, \href
  {http://adsabs.harvard.edu/abs/2006MNRAS.370..289B} {370, 289}

\bibitem[\protect\citeauthoryear{{Bouwens} et~al.,}{{Bouwens}
  et~al.}{2012}]{Bouwens:2012ApJ}
{Bouwens} R.~J.,  et~al., 2012, \mn@doi [\apj] {10.1088/0004-637X/754/2/83},
  \href {http://adsabs.harvard.edu/abs/2012ApJ...754...83B} {754, 83}

\bibitem[\protect\citeauthoryear{{Brinchmann}, {Pettini}  \&
  {Charlot}}{{Brinchmann} et~al.}{2008}]{Brinchmann:2008MNRAS}
{Brinchmann} J.,  {Pettini} M.,   {Charlot} S.,  2008, \mn@doi [\mnras]
  {10.1111/j.1365-2966.2008.12914.x}, \href
  {http://adsabs.harvard.edu/abs/2008MNRAS.385..769B} {385, 769}

\bibitem[\protect\citeauthoryear{{Bromm}, {Coppi}  \& {Larson}}{{Bromm}
  et~al.}{2002}]{Bromm:2002ApJ}
{Bromm} V.,  {Coppi} P.~S.,   {Larson} R.~B.,  2002, \mn@doi [\apj]
  {10.1086/323947}, \href {http://adsabs.harvard.edu/abs/2002ApJ...564...23B}
  {564, 23}

\bibitem[\protect\citeauthoryear{{Cai} et~al.,}{{Cai}
  et~al.}{2011}]{Cai:2011ApJ}
{Cai} Z.,  et~al., 2011, \mn@doi [\apjl] {10.1088/2041-8205/736/2/L28}, \href
  {http://adsabs.harvard.edu/abs/2011ApJ...736L..28C} {736, L28}

\bibitem[\protect\citeauthoryear{{Cassata} et~al.,}{{Cassata}
  et~al.}{2013}]{Cassata:2013A&A}
{Cassata} P.,  et~al., 2013, \mn@doi [\aap] {10.1051/0004-6361/201220969},
  \href {http://adsabs.harvard.edu/abs/2013A%26A...556A..68C} {556, A68}

\bibitem[\protect\citeauthoryear{{Daddi} et~al.,}{{Daddi}
  et~al.}{2007}]{Daddi:2007ApJ}
{Daddi} E.,  et~al., 2007, \mn@doi [\apj] {10.1086/521818}, \href
  {http://adsabs.harvard.edu/abs/2007ApJ...670..156D} {670, 156}

\bibitem[\protect\citeauthoryear{{Dayal}, {Ferrara}  \& {Gallerani}}{{Dayal}
  et~al.}{2008}]{dayal:2008mnras}
{Dayal} P.,  {Ferrara} A.,   {Gallerani} S.,  2008, \mn@doi [\mnras]
  {10.1111/j.1365-2966.2008.13721.x}, \href
  {http://adsabs.harvard.edu/abs/2008MNRAS.389.1683D} {389, 1683}

\bibitem[\protect\citeauthoryear{{De Breuck}, {R{\"o}ttgering}, {Miley}, {van
  Breugel}  \& {Best}}{{De Breuck} et~al.}{2000}]{DeBreuck:2000AA}
{De Breuck} C.,  {R{\"o}ttgering} H.,  {Miley} G.,  {van Breugel} W.,   {Best}
  P.,  2000, \aap, \href {http://adsabs.harvard.edu/abs/2000A%26A...362..519D}
  {362, 519}

\bibitem[\protect\citeauthoryear{{Elvis} et~al.}{{Elvis}
  et~al.}{2009}]{Elvis_2009}
{Elvis} M.,  et~al., 2009, \mn@doi [\apjs] {10.1088/0067-0049/184/1/158}, \href
  {http://adsabs.harvard.edu/abs/2009ApJS..184..158E} {184, 158}

\bibitem[\protect\citeauthoryear{{Erb}, {Pettini}, {Shapley}, {Steidel}, {Law}
  \& {Reddy}}{{Erb} et~al.}{2010}]{Erb:2010ApJ}
{Erb} D.~K.,  {Pettini} M.,  {Shapley} A.~E.,  {Steidel} C.~C.,  {Law} D.~R.,
  {Reddy} N.~A.,  2010, \mn@doi [\apj] {10.1088/0004-637X/719/2/1168}, \href
  {http://adsabs.harvard.edu/abs/2010ApJ...719.1168E} {719, 1168}

\bibitem[\protect\citeauthoryear{{Ferland} et~al.,}{{Ferland}
  et~al.}{2013}]{cloudy:2013}
{Ferland} G.~J.,  et~al., 2013, {Revista Mexicana de Astronomia y Astrofisica},
  \href {http://adsabs.harvard.edu/abs/2013RMxAA..49..137F} {49, 137}

\bibitem[\protect\citeauthoryear{{Ferrara}, {Salvadori}, {Yue}  \&
  {Schleicher}}{{Ferrara} et~al.}{2014}]{Ferrara_2014}
{Ferrara} A.,  {Salvadori} S.,  {Yue} B.,   {Schleicher} D.~R.~G.,  2014,
  preprint, \href {http://adsabs.harvard.edu/abs/2014arXiv1406.6685F} {}
  (\mn@eprint {arXiv} {1406.6685})

\bibitem[\protect\citeauthoryear{{Gonz{\'a}lez}, {Labb{\'e}}, {Bouwens},
  {Illingworth}, {Franx}  \& {Kriek}}{{Gonz{\'a}lez}
  et~al.}{2011}]{Gonzalez:2011}
{Gonz{\'a}lez} V.,  {Labb{\'e}} I.,  {Bouwens} R.~J.,  {Illingworth} G.,
  {Franx} M.,   {Kriek} M.,  2011, \mn@doi [\apjl]
  {10.1088/2041-8205/735/2/L34}, \href
  {http://adsabs.harvard.edu/abs/2011ApJ...735L..34G} {735, L34}

\bibitem[\protect\citeauthoryear{{Greif}, {Bromm}, {Clark}, {Glover}, {Smith},
  {Klessen}, {Yoshida}  \& {Springel}}{{Greif} et~al.}{2012}]{greif:2012mnras}
{Greif} T.~H.,  {Bromm} V.,  {Clark} P.~C.,  {Glover} S.~C.~O.,  {Smith} R.~J.,
   {Klessen} R.~S.,  {Yoshida} N.,   {Springel} V.,  2012, \mn@doi [\mnras]
  {10.1111/j.1365-2966.2012.21212.x}, \href
  {http://adsabs.harvard.edu/abs/2012MNRAS.424..399G} {424, 399}

\bibitem[\protect\citeauthoryear{{Heap}, {Bouret}  \& {Hubeny}}{{Heap}
  et~al.}{2015}]{heap:2015arxiv}
{Heap} S.,  {Bouret} J.-C.,   {Hubeny} I.,  2015, preprint, \href
  {http://adsabs.harvard.edu/abs/2015arXiv150402742H} {} (\mn@eprint {arXiv}
  {1504.02742})

\bibitem[\protect\citeauthoryear{{Kashikawa} et~al.,}{{Kashikawa}
  et~al.}{2012}]{kashikawa:2012ApJ}
{Kashikawa} N.,  et~al., 2012, \mn@doi [\apj] {10.1088/0004-637X/761/2/85},
  \href {http://adsabs.harvard.edu/abs/2012ApJ...761...85K} {761, 85}

\bibitem[\protect\citeauthoryear{{Kehrig}, {V{\'{\i}}lchez},
  {P{\'e}rez-Montero}, {Iglesias-P{\'a}ramo}, {Brinchmann}, {Kunth}, {Durret}
  \& {Bayo}}{{Kehrig} et~al.}{2015}]{kehrig:2015ApJ}
{Kehrig} C.,  {V{\'{\i}}lchez} J.~M.,  {P{\'e}rez-Montero} E.,
  {Iglesias-P{\'a}ramo} J.,  {Brinchmann} J.,  {Kunth} D.,  {Durret} F.,
  {Bayo} F.~M.,  2015, \mn@doi [\apjl] {10.1088/2041-8205/801/2/L28}, \href
  {http://adsabs.harvard.edu/abs/2015ApJ...801L..28K} {801, L28}

\bibitem[\protect\citeauthoryear{{Larson} et~al.,}{{Larson}
  et~al.}{2011}]{Larson:2011}
{Larson} D.,  et~al., 2011, \mn@doi [\apjs] {10.1088/0067-0049/192/2/16}, \href
  {http://adsabs.harvard.edu/abs/2011ApJS..192...16L} {192, 16}

\bibitem[\protect\citeauthoryear{{Latif}, {Schleicher}, {Schmidt}  \&
  {Niemeyer}}{{Latif} et~al.}{2013}]{Latif_2013c}
{Latif} M.~A.,  {Schleicher} D.~R.~G.,  {Schmidt} W.,   {Niemeyer} J.~C.,
  2013, \mn@doi [\mnras] {10.1093/mnras/stt1786}, \href
  {http://adsabs.harvard.edu/abs/2013MNRAS.436.2989L} {436, 2989}

\bibitem[\protect\citeauthoryear{{Ma}, {Maio}, {Ciardi}  \& {Salvaterra}}{{Ma}
  et~al.}{2015}]{Ma:2015MNRAS}
{Ma} Q.,  {Maio} U.,  {Ciardi} B.,   {Salvaterra} R.,  2015, \mn@doi [\mnras]
  {10.1093/mnras/stv477}, \href
  {http://adsabs.harvard.edu/abs/2015MNRAS.449.3006M} {449, 3006}

\bibitem[\protect\citeauthoryear{{Maio}, {Ciardi}, {Dolag}, {Tornatore}  \&
  {Khochfar}}{{Maio} et~al.}{2010}]{maio:2010mnras}
{Maio} U.,  {Ciardi} B.,  {Dolag} K.,  {Tornatore} L.,   {Khochfar} S.,  2010,
  \mn@doi [\mnras] {10.1111/j.1365-2966.2010.17003.x}, \href
  {http://adsabs.harvard.edu/abs/2010MNRAS.407.1003M} {407, 1003}

\bibitem[\protect\citeauthoryear{{Matthee}, {Sobral}, {Santos},
  {R{\"o}ttgering}, {Darvish}  \& {Mobasher}}{{Matthee}
  et~al.}{2015}]{Matthee:2015arXiv15}
{Matthee} J.,  {Sobral} D.,  {Santos} S.,  {R{\"o}ttgering} H.,  {Darvish} B.,
   {Mobasher} B.,  2015, \mn@doi [\mnras] {10.1093/mnras/stv947}, \href
  {http://adsabs.harvard.edu/abs/2015MNRAS.451.4919M} {451, 4919}

\bibitem[\protect\citeauthoryear{{McLure} et~al.,}{{McLure}
  et~al.}{2011}]{McLure:2011MNRAS}
{McLure} R.~J.,  et~al., 2011, \mn@doi [\mnras]
  {10.1111/j.1365-2966.2011.19626.x}, \href
  {http://adsabs.harvard.edu/abs/2011MNRAS.418.2074M} {418, 2074}

\bibitem[\protect\citeauthoryear{{Nagao} et~al.,}{{Nagao}
  et~al.}{2008}]{Nagao2008A-Photometric-S}
{Nagao} T.,  et~al., 2008, \mn@doi [\apj] {10.1086/587888}, \href
  {http://cdsads.u-strasbg.fr/abs/2008ApJ...680..100N} {680, 100}

\bibitem[\protect\citeauthoryear{{Ouchi} et~al.,}{{Ouchi}
  et~al.}{2009}]{Ouchi:2009ApJ}
{Ouchi} M.,  et~al., 2009, \mn@doi [\apj] {10.1088/0004-637X/696/2/1164}, \href
  {http://adsabs.harvard.edu/abs/2009ApJ...696.1164O} {696, 1164}

\bibitem[\protect\citeauthoryear{{Pacucci} \& {Ferrara}}{{Pacucci} \&
  {Ferrara}}{2015}]{Pacucci_2015}
{Pacucci} F.,  {Ferrara} A.,  2015, \mn@doi [\mnras] {10.1093/mnras/stv018},
  \href {http://adsabs.harvard.edu/abs/2015MNRAS.448..104P} {448, 104}

\bibitem[\protect\citeauthoryear{{Pacucci}, {Ferrara}, {Volonteri}  \&
  {Dubus}}{{Pacucci} et~al.}{2015}]{pacucci_2015_spettro}
{Pacucci} F.,  {Ferrara} A.,  {Volonteri} M.,   {Dubus} G.,  2015, preprint,
  \href {http://adsabs.harvard.edu/abs/2015arXiv150605299P} {} (\mn@eprint
  {arXiv} {1506.05299})

\bibitem[\protect\citeauthoryear{{Pallottini}, {Ferrara}, {Gallerani},
  {Salvadori}  \& {D'Odorico}}{{Pallottini}
  et~al.}{2014a}]{pallottini:2014_sim}
{Pallottini} A.,  {Ferrara} A.,  {Gallerani} S.,  {Salvadori} S.,   {D'Odorico}
  V.,  2014a, \mn@doi [\mnras] {10.1093/mnras/stu451}, \href
  {http://adsabs.harvard.edu/abs/2014MNRAS.440.2498P} {440, 2498}

\bibitem[\protect\citeauthoryear{{Pallottini}, {Gallerani}  \&
  {Ferrara}}{{Pallottini} et~al.}{2014b}]{pallottini:2014_cgmh}
{Pallottini} A.,  {Gallerani} S.,   {Ferrara} A.,  2014b, \mn@doi [\mnras]
  {10.1093/mnrasl/slu126}, \href
  {http://adsabs.harvard.edu/abs/2014MNRAS.444L.105P} {444, L105}

\bibitem[\protect\citeauthoryear{{Pallottini}, {Gallerani}, {Ferrara}, {Yue},
  {Vallini}, {Maiolino}  \& {Feruglio}}{{Pallottini}
  et~al.}{2015}]{pallottinicmb}
{Pallottini} A.,  {Gallerani} S.,  {Ferrara} A.,  {Yue} B.,  {Vallini} L.,
  {Maiolino} R.,   {Feruglio} C.,  2015, preprint, \href
  {http://adsabs.harvard.edu/abs/2015arXiv150605803P} {} (\mn@eprint {arXiv}
  {1506.05803})

\bibitem[\protect\citeauthoryear{{Raiter}, {Schaerer}  \& {Fosbury}}{{Raiter}
  et~al.}{2010}]{Raiter:2010A&A}
{Raiter} A.,  {Schaerer} D.,   {Fosbury} R.~A.~E.,  2010, \mn@doi [\aap]
  {10.1051/0004-6361/201015236}, \href
  {http://adsabs.harvard.edu/abs/2010A%26A...523A..64R} {523, A64}

\bibitem[\protect\citeauthoryear{{Regan}, {Johansson}  \& {Wise}}{{Regan}
  et~al.}{2014}]{Regan_2014}
{Regan} J.~A.,  {Johansson} P.~H.,   {Wise} J.~H.,  2014, \mn@doi [\apj]
  {10.1088/0004-637X/795/2/137}, \href
  {http://adsabs.harvard.edu/abs/2014ApJ...795..137R} {795, 137}

\bibitem[\protect\citeauthoryear{{Salvadori} \& {Ferrara}}{{Salvadori} \&
  {Ferrara}}{2009}]{salvadori:2009mnras}
{Salvadori} S.,  {Ferrara} A.,  2009, \mn@doi [\mnras]
  {10.1111/j.1745-3933.2009.00627.x}, \href
  {http://adsabs.harvard.edu/abs/2009MNRAS.395L...6S} {395, L6}

\bibitem[\protect\citeauthoryear{{Salvadori}, {Tolstoy}, {Ferrara}  \&
  {Zaroubi}}{{Salvadori} et~al.}{2014}]{salvadori:2014mnras}
{Salvadori} S.,  {Tolstoy} E.,  {Ferrara} A.,   {Zaroubi} S.,  2014, \mn@doi
  [\mnras] {10.1093/mnrasl/slt132}, \href
  {http://adsabs.harvard.edu/abs/2014MNRAS.437L..26S} {437, L26}

\bibitem[\protect\citeauthoryear{{Schaerer}}{{Schaerer}}{2002}]{Schaerer:2002A&A}
{Schaerer} D.,  2002, \mn@doi [\aap] {10.1051/0004-6361:20011619}, \href
  {http://adsabs.harvard.edu/abs/2002A%26A...382...28S} {382, 28}

\bibitem[\protect\citeauthoryear{{Schneider}, {Ferrara}, {Natarajan}  \&
  {Omukai}}{{Schneider} et~al.}{2002}]{schneider:2002apj}
{Schneider} R.,  {Ferrara} A.,  {Natarajan} P.,   {Omukai} K.,  2002, \mn@doi
  [\apj] {10.1086/339917}, \href
  {http://adsabs.harvard.edu/abs/2002ApJ...571...30S} {571, 30}

\bibitem[\protect\citeauthoryear{{Schneider}, {Omukai}, {Inoue}  \&
  {Ferrara}}{{Schneider} et~al.}{2006}]{schneider:2006mnras}
{Schneider} R.,  {Omukai} K.,  {Inoue} A.~K.,   {Ferrara} A.,  2006, \mn@doi
  [\mnras] {10.1111/j.1365-2966.2006.10391.x}, \href
  {http://adsabs.harvard.edu/abs/2006MNRAS.369.1437S} {369, 1437}

\bibitem[\protect\citeauthoryear{{Shang}, {Bryan}  \& {Haiman}}{{Shang}
  et~al.}{2010}]{Shang_2010}
{Shang} C.,  {Bryan} G.~L.,   {Haiman} Z.,  2010, \mn@doi [\mnras]
  {10.1111/j.1365-2966.2009.15960.x}, \href
  {http://adsabs.harvard.edu/abs/2010MNRAS.402.1249S} {402, 1249}

\bibitem[\protect\citeauthoryear{{Sheth} \& {Tormen}}{{Sheth} \&
  {Tormen}}{1999}]{sheth:1999MNRAS}
{Sheth} R.~K.,  {Tormen} G.,  1999, \mn@doi [\mnras]
  {10.1046/j.1365-8711.1999.02692.x}, \href
  {http://adsabs.harvard.edu/abs/1999MNRAS.308..119S} {308, 119}

\bibitem[\protect\citeauthoryear{{Sobral}, {Matthee}, {Darvish}, {Schaerer},
  {Mobasher}, {R{\"o}ttgering}, {Santos}  \& {Hemmati}}{{Sobral}
  et~al.}{2015}]{Sobral:2015arXiv15}
{Sobral} D.,  {Matthee} J.,  {Darvish} B.,  {Schaerer} D.,  {Mobasher} B.,
  {R{\"o}ttgering} H.,  {Santos} S.,   {Hemmati} S.,  2015, \apj, \href
  {http://adsabs.harvard.edu/abs/2015arXiv150401734S} {808, 139}

\bibitem[\protect\citeauthoryear{{Stark}, {Schenker}, {Ellis}, {Robertson},
  {McLure}  \& {Dunlop}}{{Stark} et~al.}{2013}]{Stark:2013ApJ}
{Stark} D.~P.,  {Schenker} M.~A.,  {Ellis} R.,  {Robertson} B.,  {McLure} R.,
  {Dunlop} J.,  2013, \mn@doi [\apj] {10.1088/0004-637X/763/2/129}, \href
  {http://adsabs.harvard.edu/abs/2013ApJ...763..129S} {763, 129}

\bibitem[\protect\citeauthoryear{{Sugimura}, {Omukai}  \& {Inoue}}{{Sugimura}
  et~al.}{2014}]{Sugimura_2014}
{Sugimura} K.,  {Omukai} K.,   {Inoue} A.~K.,  2014, \mn@doi [\mnras]
  {10.1093/mnras/stu1778}, \href
  {http://adsabs.harvard.edu/abs/2014MNRAS.445..544S} {445, 544}

\bibitem[\protect\citeauthoryear{{Teyssier}}{{Teyssier}}{2002}]{teyssier:2002}
{Teyssier} R.,  2002, \mn@doi [\aap] {10.1051/0004-6361:20011817}, \href
  {http://adsabs.harvard.edu/abs/2002A%26A...385..337T} {385, 337}

\bibitem[\protect\citeauthoryear{{Tornatore}, {Ferrara}  \&
  {Schneider}}{{Tornatore} et~al.}{2007}]{Tornatore:2007MNRAS}
{Tornatore} L.,  {Ferrara} A.,   {Schneider} R.,  2007, \mn@doi [\mnras]
  {10.1111/j.1365-2966.2007.12215.x}, \href
  {http://adsabs.harvard.edu/abs/2007MNRAS.382..945T} {382, 945}

\bibitem[\protect\citeauthoryear{{Trenti}, {Stiavelli}  \& {Michael
  Shull}}{{Trenti} et~al.}{2009}]{trenti:2009b}
{Trenti} M.,  {Stiavelli} M.,   {Michael Shull} J.,  2009, \mn@doi [\apj]
  {10.1088/0004-637X/700/2/1672}, \href
  {http://adsabs.harvard.edu/abs/2009ApJ...700.1672T} {700, 1672}

\bibitem[\protect\citeauthoryear{{Turk}, {Abel}  \& {O'Shea}}{{Turk}
  et~al.}{2009}]{Turk:2009Sci}
{Turk} M.~J.,  {Abel} T.,   {O'Shea} B.,  2009, \mn@doi [Science]
  {10.1126/science.1173540}, \href
  {http://adsabs.harvard.edu/abs/2009Sci...325..601T} {325, 601}

\bibitem[\protect\citeauthoryear{{Visbal}, {Haiman}  \& {Bryan}}{{Visbal}
  et~al.}{2015}]{Visbal:2015arXiv}
{Visbal} E.,  {Haiman} Z.,   {Bryan} G.~L.,  2015, preprint, \href
  {http://adsabs.harvard.edu/abs/2015arXiv150506359V} {} (\mn@eprint {arXiv}
  {1505.06359})

\bibitem[\protect\citeauthoryear{{Volonteri}, {Lodato}  \&
  {Natarajan}}{{Volonteri} et~al.}{2008}]{Volonteri_2008}
{Volonteri} M.,  {Lodato} G.,   {Natarajan} P.,  2008, \mn@doi [\mnras]
  {10.1111/j.1365-2966.2007.12589.x}, \href
  {http://adsabs.harvard.edu/abs/2008MNRAS.383.1079V} {383, 1079}

\bibitem[\protect\citeauthoryear{{Wise}, {Turk}, {Norman}  \& {Abel}}{{Wise}
  et~al.}{2012}]{wise:2012apj}
{Wise} J.~H.,  {Turk} M.~J.,  {Norman} M.~L.,   {Abel} T.,  2012, \mn@doi
  [\apj] {10.1088/0004-637X/745/1/50}, \href
  {http://adsabs.harvard.edu/abs/2012ApJ...745...50W} {745, 50}

\bibitem[\protect\citeauthoryear{{Xu}, {Wise}  \& {Norman}}{{Xu}
  et~al.}{2013}]{xu:2013arxiv}
{Xu} H.,  {Wise} J.~H.,   {Norman} M.~L.,  2013, \mn@doi [\apj]
  {10.1088/0004-637X/773/2/83}, \href
  {http://adsabs.harvard.edu/abs/2013ApJ...773...83X} {773, 83}

\bibitem[\protect\citeauthoryear{{Yoshida}, {Omukai}, {Hernquist}  \&
  {Abel}}{{Yoshida} et~al.}{2006}]{Yoshida:2006ApJ}
{Yoshida} N.,  {Omukai} K.,  {Hernquist} L.,   {Abel} T.,  2006, \mn@doi [\apj]
  {10.1086/507978}, \href {http://adsabs.harvard.edu/abs/2006ApJ...652....6Y}
  {652, 6}

\bibitem[\protect\citeauthoryear{{Yue}, {Ferrara}, {Salvaterra}, {Xu}  \&
  {Chen}}{{Yue} et~al.}{2013}]{Yue_2013}
{Yue} B.,  {Ferrara} A.,  {Salvaterra} R.,  {Xu} Y.,   {Chen} X.,  2013,
  \mn@doi [\mnras] {10.1093/mnras/stt826}, \href
  {http://adsabs.harvard.edu/abs/2013MNRAS.433.1556Y} {433, 1556}

\bibitem[\protect\citeauthoryear{{Yue}, {Ferrara}, {Salvaterra}, {Xu}  \&
  {Chen}}{{Yue} et~al.}{2014}]{yue:2014mnras}
{Yue} B.,  {Ferrara} A.,  {Salvaterra} R.,  {Xu} Y.,   {Chen} X.,  2014,
  \mn@doi [\mnras] {10.1093/mnras/stu351}, \href
  {http://adsabs.harvard.edu/abs/2014MNRAS.440.1263Y} {440, 1263}

\bibitem[\protect\citeauthoryear{{Zabl}, {N{\o}rgaard-Nielsen}, {Fynbo},
  {Laursen}, {Ouchi}  \& {Kj{\ae}rgaard}}{{Zabl}
  et~al.}{2015}]{Zabl:2015arXiv15}
{Zabl} J.,  {N{\o}rgaard-Nielsen} H.~U.,  {Fynbo} J.~P.~U.,  {Laursen} P.,
  {Ouchi} M.,   {Kj{\ae}rgaard} P.,  2015, \mn@doi [\mnras]
  {10.1093/mnras/stv1019}, \href
  {http://adsabs.harvard.edu/abs/2015MNRAS.451.2050Z} {451, 2050}

\bibitem[\protect\citeauthoryear{{Zheng} et~al.,}{{Zheng}
  et~al.}{2012}]{Zheng:2012Natur}
{Zheng} W.,  et~al., 2012, \mn@doi [\nat] {10.1038/nature11446}, \href
  {http://adsabs.harvard.edu/abs/2012Natur.489..406Z} {489, 406}

\makeatother
\end{thebibliography}
\bsp

\label{lastpage}

\end{document}